\documentclass[aps,pra,twocolumn,preprintnumbers]{revtex4}
\usepackage{amsmath}
\usepackage{amsmath,amssymb,bm,graphicx,subfigure}

\begin{document}

\title{Conditions of Low Dimensionality for Strongly Interacting Atoms Under a
Transverse Trap}
\author{J. P. Kestner, L.-M. Duan}
\address{FOCUS center and MCTP, Department of Physics, University of Michigan, Ann
Arbor, MI 48109}

\begin{abstract}
For a dilute atomic gas in a strong transverse trapping potential, one
normally expects that, in the ground state, the gas will populate only the
lowest transverse level. We show, however, that for the strongly interacting
gas under a Feshbach resonance, the ground state includes a large fraction
of atoms in excited levels of the trap, even if the gas is very dilute and
the trap is very strong. This is because the effective atom-molecule
coupling is typically enhanced to many times the trap mode spacing by an
induced confinement along the untrapped dimension(s). Thus one cannot
''freeze out'' the transverse degrees of freedom except under certain
conditions.
\end{abstract}

\maketitle

\affiliation{Department of Physics, University of Michigan, Ann
Arbor, MI 48109}

\section{Introduction}

Recently, the system of interacting atomic gas at low dimensions has
attracted considerable interest as it supports a wealth of physics
\cite{1,2,3,4,5,6,7}. For ultracold atoms, low dimensions are
typically achieved by applying a transverse optical lattice
potential \cite{8}. With a one (or two) dimensional deep optical
lattice potential, the atoms trapped at different potential wells
basically do not interact with each other. For weakly interacting
gas, one can then freeze out the transverse degrees of freedom by
assuming the system to be in the lowest level of the trap potential.
The atomic gas at each individual potential well thus behaves as an
effective two (or one) dimensional system.

The condition for achieving effective low-dimensionality in an
optical lattice becomes more involved if one has strongly
interacting atomic gas. One of the most exciting directions in
current atomic physics is to study this strongly interacting
atomic gas, where the effective interaction as measured by the
atomic scattering length can be tuned in the full range via a
Feshbach resonance through control of an external magnetic field
\cite{9}. The Feshbach resonance basically comes from the coupling
of atoms in the open collision channel to the molecules in the
closed channel. If this atom-molecule coupling rate becomes larger
than the level spacing of the transverse trapping potential, one
cannot assume a low-dimensional system by freezing the transverse
mode in the lowest level \cite{10,11}. On the other hand,
intuitively, one may expect that the effective atom-molecule
coupling rate always decreases when we make the gas more dilute,
and the transverse level spacing increases when we raise the
trapping laser intensity. So, for a sufficiently dilute gas under
a strong transverse trap, one can still get the transverse level
spacing larger than the atom-molecule coupling rate. With this
expectation, several recent works have studied properties of
low-dimensional strongly interacting gas by (implicitly) assuming
the transverse mode in the trap ground state \cite{5,6}.

In this work, we want to show that the condition of low dimensionality for
strongly interacting atoms under a transverse trap is more subtle than the
above simple picture. We will show that even if the gas becomes very dilute
(in the limiting case, one can just have two atoms with the gas density
tending to zero), for the ground state of the system, we will still have a
significant (actually, dominant) fraction of atomic population in the
excited transverse levels under a typical wide Feshbach resonance.
Furthermore, it is very ineffective to reduce the transverse excited
fraction by increasing the trapping laser intensity. For realistic atoms
such as $^{40}$K or $^{6}$Li, even if the trapping potential is increased to
some completely impractical level, the transverse excited fraction is not
yet negligible. The basic reason for this unusual phenomena is that in the
low-dimensional trap, there is an increased tendency for atoms to pair up
spatially also along the untrapped dimension(s). This effect lends itself to
an intuitive understanding of the coupling enhancement in the dilute limit.
Furthermore, with stronger transverse traps, the atoms in the induced pairs
will become more spatially confined also along the untrapped dimension(s).
So, although the transverse level spacing increases, the effective
atom-molecule coupling is also significantly enhanced. As a net effect, the
atomic population in the transverse excited levels becomes pretty
insensitive to the magnitude of the trapping potential. Although the result
here does not preclude the possibility of an effective low-dimensional
Hamiltonian for strongly interacting atoms under the transverse trap, it
indeed shows that in general, one can not neglect the atomic population in
the transverse excited levels, and derivation of an effective
low-dimensional Hamiltonian would be more subtle and tricky than one naively
expects \cite{10,12}.

In the next section, we give the general formalism for strongly
interacting atoms under a transverse trap in the dilute gas limit.
In that limit, the basic picture is captured by two-body physics.
For the ground state of the system, the atoms form pairs, and
interaction between the pairs become negligible. There have been a
few works on description of two-body physics of strongly
interacting atoms in a trap using the single-channel model \cite
{7}. Recently, there have also been descriptions of the problem
with a more realistic two-channel model for the Feshbach resonance
(in one-dimensional or three-dimensional traps \cite{11,13}). The
formalism there, however, neglects the atomic background
scattering. We extend this formalism to include the background
scattering, which becomes necessary when the system is outside of
the near-resonance region. In Sec. III, we present our main
calculation results for the condition of low-dimensionality under
a transverse one-dimensional or two-dimensional traps. We will
also give detailed studies of the atomic population distributions
in the transverse levels and in the free dimension(s) for the
$^{40}$K and $^{6}$Li atoms under trapping potentials of various
intensities.

\section{General formalism}

To achieve a $d$-dimensional ($d<3$) atomic gas, we assume a $(3-d)$%
-dimensional optical lattice applied along the transverse direction.
The lattice potential barrier is high so that the atomic interaction
between different potential wells becomes negligible. With a strong
lattice potential, the atoms at the bottom of the potential wells
basically see a transverse harmonic trap. The atomic gas in each
well can then be modeled as in a $\left( 3-d\right) $-dimensional
harmonic trap of a characteristic trapping frequency $\omega $. The
atoms are of mass $m$ and possess internal states $\sigma
=\{\uparrow ,\downarrow \}$. We treat the problem of strongly
interacting atoms across a Feshbach resonance using the standard
two-channel model \cite{14}. The short range interaction between
closed-channel molecules and open-channel atoms is modeled by a
delta function. The Hamiltonian is expressed as $H=H_{0}+H_{I}$,
with
\begin{multline}
H_{0}=\sum_{\sigma =\uparrow ,\downarrow }\int d^{3}\mathbf{r}\Psi _{\sigma
}^{\dagger }\left( -\frac{\hbar ^{2}\nabla ^{2}}{2m}+\frac{1}{2}m\omega
^{2}\sum_{i=1}^{3-d}x_{i}^{2}\right) \Psi _{\sigma }  \notag \\
+\int d^{3}\mathbf{r}\Phi ^{\dagger }\left( -\frac{\hbar ^{2}\nabla ^{2}}{4m}%
+m\omega ^{2}\sum_{i=1}^{3-d}x_{i}^{2}+\bar{\nu}_{b}\right) \Phi
\end{multline}
and
\begin{equation}
H_{I}=\bar{g}_{b}\int d^{3}\mathbf{r}\left( \Psi _{\uparrow }^{\dagger }\Psi
_{\downarrow }^{\dagger }\Phi +h.c.\right) +\bar{U}_{b}\int d^{3}\mathbf{r}%
\Psi _{\uparrow }^{\dagger }\Psi _{\downarrow }^{\dagger }\Psi _{\downarrow
}\Psi _{\uparrow }\;,
\end{equation}
where $\Psi \left( \mathbf{r}\right) $ is the atomic field operator, $\Phi
\left( \mathbf{r}\right) $ is the molecular field operator, $\bar{\nu}_{b}$
is the bare detuning, $\bar{g}_{b}$ is the bare atom-molecule coupling rate,
and $\bar{U}_{b}$ is the bare background atomic scattering rate. The bare
parameters are related to the physical parameters via the standard
renormalization relations \cite{15}:
\begin{gather}
\bar{U}_{c}^{-1}=-\int \frac{d^{3}\mathbf{k}}{\left( 2\pi \right) ^{3}}\frac{%
1}{2\epsilon _{\mathbf{k}}}\,,\quad \Gamma ^{-1}=1+\frac{\bar{U}_{p}}{U_{c}}%
\,  \notag  \label{eq:renormalization} \\
\bar{U}_{b}=\Gamma \bar{U}_{p}\,,\quad \bar{g}_{b}=\Gamma \bar{g}%
_{p}\,,\quad \bar{\nu}_{p}=\bar{\nu}_{b}+\Gamma \frac{\bar{g}_{p}^{2}}{%
\bar{U}_{c}}\,
\end{gather}
where the subscript $p$ denotes physical parameters, $\epsilon _{\mathbf{k}%
}=\hbar ^{2}\mathbf{k}^{2}/2m$, and the integral is taken in three
dimensions with an explicit energy cutoff $E_{c}$ imposed on two
dimensions \cite{Note}. Then
$\bar{U}_{c}^{-1}=m^{3/2}\sqrt{E_{c}}/2^{3/2}\pi \hbar ^{3} $. The
physical parameters $\bar{g}_{p},\bar{U}_{p},\bar{\nu}_{p}$ are
determined from the scattering data as $\bar{U}_{p}=4\pi \hbar ^{2}a_{bg}/m$%
, $\bar{g}_{p}=\sqrt{4\pi \hbar ^{2}\mu _{co}W|a_{bg}|/m}$, and $\bar{\nu}%
_{p}=\mu _{co}(B-B_{0})$ ($\mu _{co}$ is the difference in magnetic
moments between the two channels) \cite{10}, where we have assumed
that the s-wave
scattering length near resonance has the form $a_{s}=a_{bg}\left( 1-\frac{%
W}{B-B_{0}}\right) $, with $a_{bg}$ as the background scattering
length, $W$ as the resonance width, and $B_{0}$ as the resonance
point.

We assume the gas to be sufficiently dilute so that we need consider only
two-body physics within each potential well. At a very low temperature, two
atoms interact and form bound atom-pairs. The interaction between the atom
pairs is negligible in the limit of a very dilute gas. The essential physics
is then captured by considering the state of two atoms under the above
interaction Hamiltonian. For the two-body physics, the center-of-mass degree
of freedom is not influenced by the interaction and can be separated under a
harmonic potential. So we can assume the center-of-mass mode is in the
ground state of the transverse trap and has zero momentum in the free
dimension(s). In this center-of-mass frame, expanding the field operators $%
\Psi \left( \mathbf{r}\right) $ and $\Phi \left( \mathbf{r}\right) $ in Eq.
(1) in terms of harmonic modes in the trapped dimensions and plane waves in
the untrapped dimensions yields
\begin{equation}
H_{0}=\hbar \omega \sum_{\mathbf{mk}\sigma }\epsilon _{\mathbf{mk}}a_{%
\mathbf{mk}\sigma }^{\dagger }a_{\mathbf{mk}\sigma }+\bar{\nu}_{b}b^{\dagger
}b\;,
\end{equation}
where $\mathbf{m}$ indexes trap eigenmodes $\{m_{i}\}$, $i=1,\ldots ,3-d$,
and $\mathbf{k}$ denotes the wave vector in the untrapped dimensions $%
\{k_{j}\}$, $j=1,\ldots ,d$. The operators $a_{\mathbf{mk}\sigma }$
and $b$ represent the corresponding atomic and molecular modes,
respectively. As there is only one molecular mode in the
center-of-mass frame, we drop the index of $b$. We have excluded the
center-of-mass energy in the Hamiltonian (4). The atom relative
energy $\epsilon _{\mathbf{mk}}$ is given (in the unit of $\hbar
\omega $) by
\begin{equation}
\epsilon _{\mathbf{mk}}=\frac{3-d}{4}+\sum_{i=1}^{3-d}m_{i}+\frac{a_{t}^{2}}{%
2}\sum_{j=1}^{d}k_{j}^{2}\;,
\end{equation}
where $a_{t}=\sqrt{\hbar /m\omega }$ characterizes the trap length scale.
Likewise, the interaction Hamiltonian $H_{I}$ in terms of these modes has
the form
\begin{multline}
H_{I}=\frac{\bar{g}_{b}}{a_{t}^{\left( 3-d\right) /2}L^{d/2}}\sum_{\mathbf{%
mnk}}\gamma _{\mathbf{mn}}\left( a_{\mathbf{mk}\uparrow }^{\dagger }a_{%
\mathbf{n-k}\downarrow }^{\dagger }b+h.c.\right) \\
+\frac{\bar{U}_{b}}{a_{t}^{\left( 3-d\right) }L^{d}}\sum_{\substack{ \mathbf{%
mnk}  \\ \mathbf{m^{\prime }n^{\prime }k^{\prime }}}}\gamma _{\mathbf{mn}%
}\gamma _{\mathbf{m^{\prime }n^{\prime }}}a_{\mathbf{mk}\uparrow }^{\dagger
}a_{\mathbf{n-k}\downarrow }^{\dagger }a_{\mathbf{n^{\prime }-k^{\prime }}%
\downarrow }a_{\mathbf{m^{\prime }k^{\prime }}\uparrow }\;,
\end{multline}
where
\begin{equation}
\gamma _{\mathbf{mn}}=\prod_{j=1}^{3-d}
\begin{cases}
\frac{\left( -1\right) ^{\left( m_{j}-n_{j}\right) /2}}{\left( 2\pi
^{3}\right) ^{1/4}\sqrt{m_{j}!n_{j}!}}\,\Gamma \left( \frac{m_{j}+n_{j}+1}{2}%
\right) & m_{j}+n_{j}\;\text{even} \\
0 & m_{j}+n_{j}\;\text{odd}
\end{cases}
.
\end{equation}
For convenience, we define dimensionless bare parameters
\begin{equation}\label{eq:dimensionless_params}
g_{b}=\frac{\bar{g}_{b}a_{t}^{-3/2}}{\hbar \omega }\,,\quad U_{b}=\frac{\bar{%
U}_{b}a_{t}^{-3}}{\hbar \omega }\,,\quad \nu
_{b}=\bar{\nu}_{b}/\hbar \omega \,
\end{equation}
and likewise for dimensionless physical parameters $g_{p}$, $U_{p}$, and $%
\nu _{p}$.

A general two-body state for the atoms and the molecule can be
expressed as \cite{11}
\begin{equation}
|\Psi \rangle =\left( \beta b^{\dagger }+\sum_{\mathbf{mnk}}\eta _{\mathbf{%
mnk}}a_{\mathbf{mk}\uparrow }^{\dagger }a_{\mathbf{n-k}\downarrow }^{\dagger
}\right) |vac\rangle .
\end{equation}
The coefficients in this ansatz state (normalized to unity) are
obtained by solving the Schr\"{o}dinger equation $H|\Psi \rangle
=E\hbar \omega |\Psi \rangle $, which yields
\begin{align}
\frac{1}{U_{b}^{eff}\left( E\right) }& =S\left( E\right) ,
\label{eq:energy} \\
\beta ^{-2}& =1-Z_b^{2}\left( E\right) \frac{\partial S\left( E\right) }{%
\partial E}, \label{eq:boson}\\
\eta _{\mathbf{mnk}}& =\beta \gamma _{\mathbf{mn}}\left( a_{t}/L\right)
^{d/2}\frac{Z_b\left( E\right) }{E-\epsilon _{\mathbf{mk}}-\epsilon _{\mathbf{%
nk}}}. \label{eq:fermion}
\end{align}
where
\begin{equation}\label{eq:U_eff}
U_{b}^{eff}\left( E\right) \equiv U_{b}-\frac{g_{b}^{2}}{\nu _{b}-E}%
\;,\;Z_{b}\left( E\right) \equiv g_{b}-\frac{U_{b}}{g_{b}}\left( \nu
_{b}-E\right) \,,
\end{equation}
and
\begin{equation}\label{eq:S}
S\left( E\right) \equiv \left( \frac{a_{t}}{L}\right) ^{d}\sum_{\mathbf{mnk}}%
\frac{\gamma _{\mathbf{mn}}^{2}}{E-\epsilon _{\mathbf{mk}}-\epsilon _{%
\mathbf{nk}}}\;.
\end{equation}

Thus, Eq. \eqref{eq:energy} determines the eigenenergy $E$, and
Eqs. \eqref{eq:boson} and \eqref{eq:fermion} give us the
eigenstate as a function of the eigenenergy. These equations are
expressed in terms of the bare parameters, and we need to the use
the renormalization relation (2) to transfer them into the ones
with the physical parameters. One can easily
check that under the relation (2), $Z_{p}\left( E\right) \equiv g_{p}-\frac{%
U_{p}}{g_{p}}\left( \nu _{p}-E\right) =Z_{b}\left( E\right) $ for any $E$,
and $\left[ U_{p}^{eff}\left( E\right) \right] ^{-1}\equiv \left( U_{p}-%
\frac{g_{p}^{2}}{\nu _{p}-E}\right) ^{-1}=\left[ U_{b}^{eff}\left( E\right) %
\right] ^{-1}-\bar{U}_{c}^{-1}a_{t}^{3}\hbar \omega $. The divergence in $%
\bar{U}_{c}^{-1}$ then exactly cancels with the divergence in $S\left(
E\right) $. As a net result, the above equations \eqref{eq:energy}--%
\eqref{eq:U_eff} retain the same form upon renormalization --- all
the bare parameters are replaced by their physical counterparts, and
$S\left( E\right) $ and $\partial S\left( E\right) /\partial E$ are
replaced by
\begin{align}
S_{p}\left( E\right) & =\frac{-1}{2^{5/2}\pi }
\begin{cases}
\zeta \left( 1/2,1/2-E/2\right)  & d=1 \\
\int_{0}^{\infty }ds\left( \frac{\Gamma \left( s+\frac{1}{4}-\frac{E}{2}%
\right) }{\Gamma \left( s+\frac{3}{4}-\frac{E}{2}\right) }-\frac{1}{\sqrt{s}}%
\right)  & d=2
\end{cases}
\\
\frac{\partial S_{p}\left( E\right) }{\partial E}& =\frac{-1}{2^{7/2}\pi }
\begin{cases}
\frac{1}{2}\zeta \left( 3/2,1/2-E/2\right)  & \quad d=1 \\
\frac{\Gamma \left( 1/4-E/2\right) }{\Gamma \left( 3/4-E/2\right) } & \quad
d=2
\end{cases}
\end{align}
where $\zeta \left( s,x\right) =\lim_{N\rightarrow \infty
}\sum_{n=0}^{N}\left( n+x\right) ^{-s}-\frac{\left( N+x\right) ^{-s+1}}{-s+1}
$ is the Hurwitz zeta function and $\Gamma \left( x\right) $ is the gamma
function. The above set of equations serve as the basic formalism for
determining the state of two atoms in a transverse trap across a Feshbach
resonance. If we take the one-dimensional case ($d=1$) and neglect the
background scattering (let $U_{p}=0$), the above equation for $E$ is reduced
to the energy equation in Ref. \cite{13}, where it is derived with a
different renormalization method.

\section{Transverse population distributions for $^{40}$K and $^{6}$Li and
conditions for low-dimensionality}

In this section, we answer the following basic question: For a
dilute gas under a strong transverse trap, can we assume the atoms
only populate the lowest transverse level at low temperature so that
the system becomes effectively low-dimensional? In the extreme limit
of a dilute gas, we can consider only two atoms in each potential
well (the density in the free dimension(s) tends to zero in this
case). We will show that even in this extreme limit the excited
fraction is typically still significant (dominant) for realistic
atoms. If the gas density becomes higher, the excited fraction
surely cannot decrease.

\begin{figure*}[ht]
\subfigure[$^{40}$K]
        {\label{fig:EvsB_K}\includegraphics[width=.6\columnwidth]
        {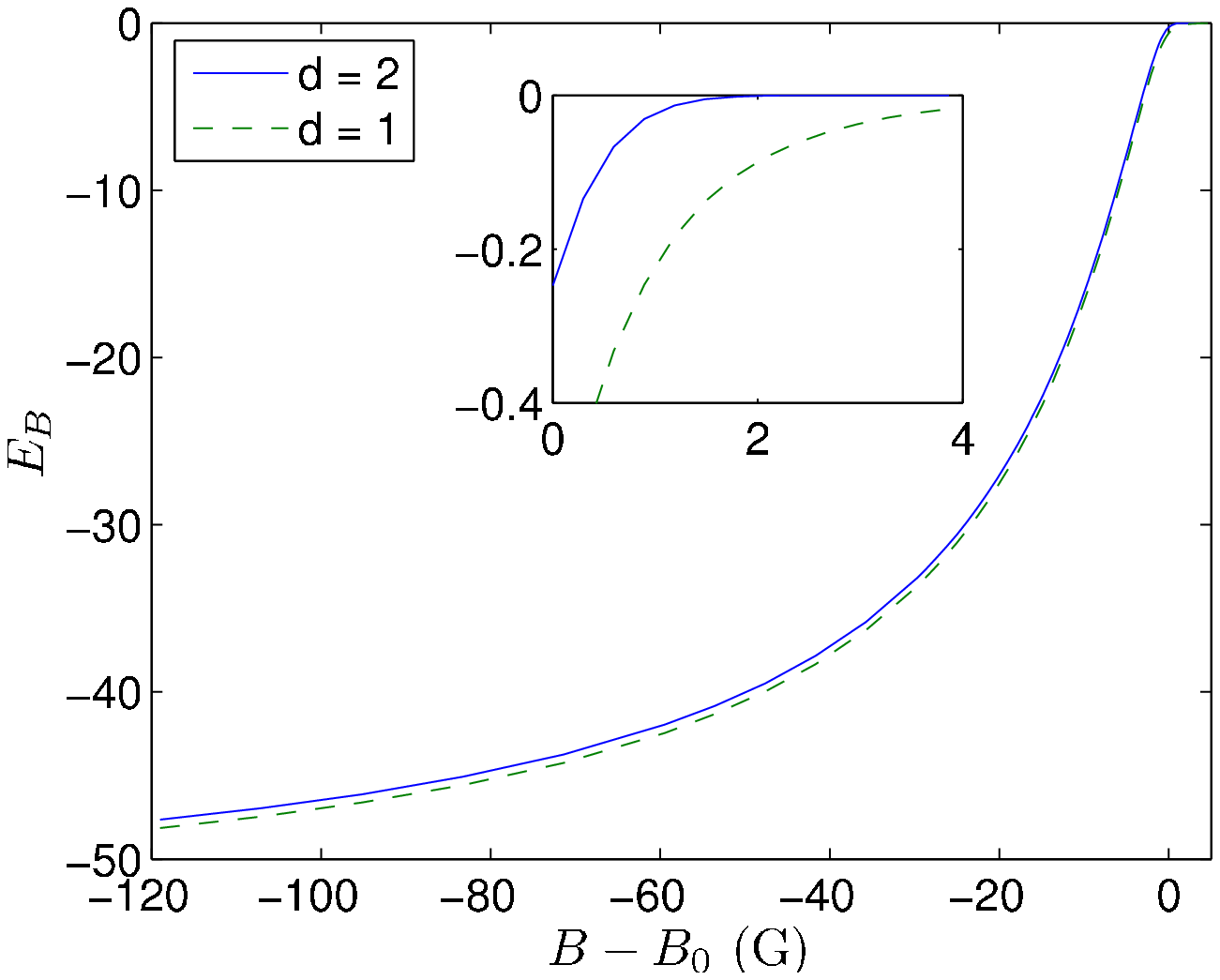}}
\subfigure[$^6$Li]
        {\label{fig:EvsB_L}\includegraphics[width=.62\columnwidth]
        {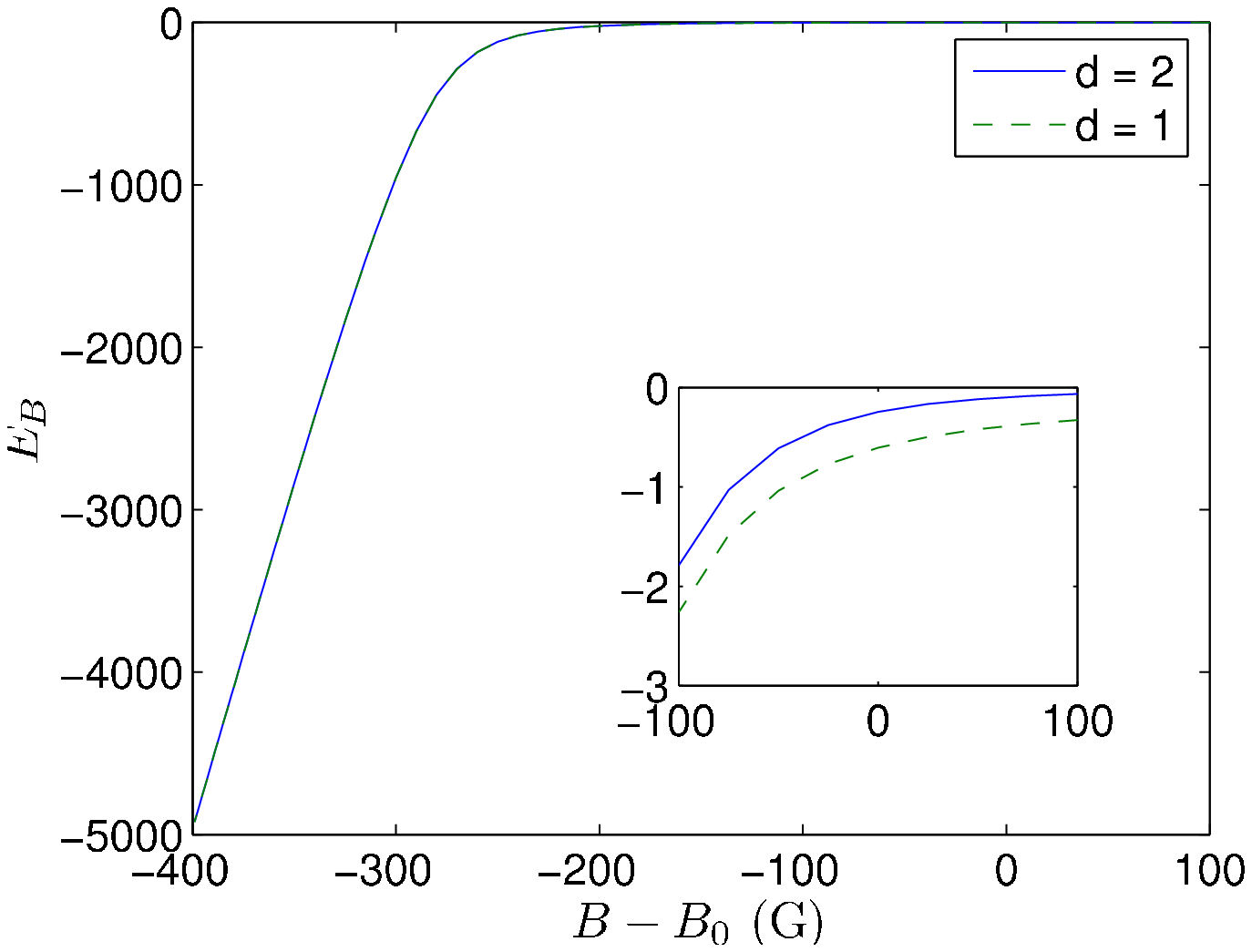}}
\caption{Binding energy vs detuning at $\protect\omega \simeq
2\protect\pi \times 62\,kHz$. $B_{0}$ is the resonant point in the
absence of an optical lattice. The inserts show a close-up of the
binding energy in the near-resonance region.} \label{fig:EvsB}
\subfigure[2D free $^{40}$K]
        {\includegraphics[width=.73\columnwidth]{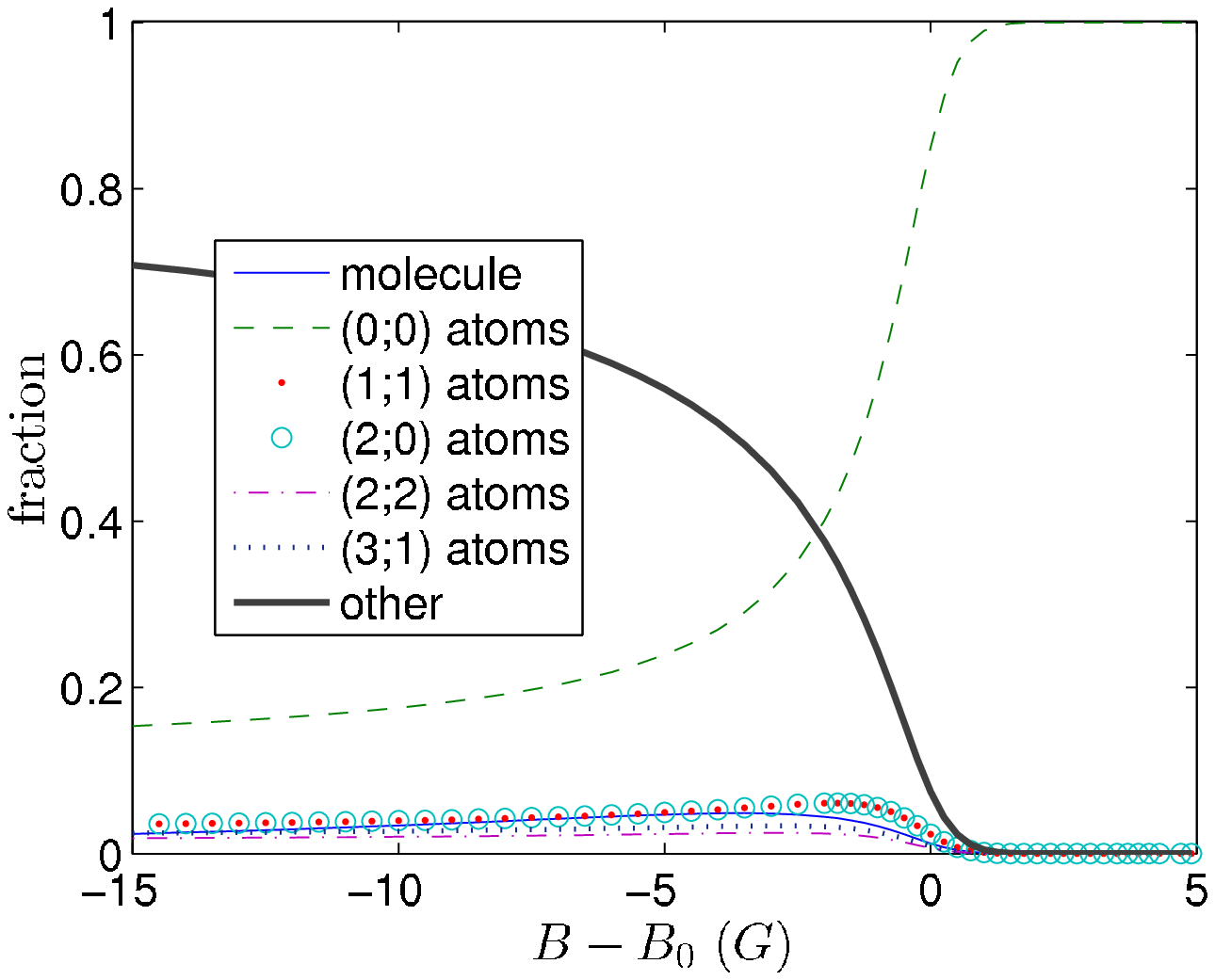}}
\subfigure[1D free $^{40}$K]
        {\includegraphics[width=.73\columnwidth]{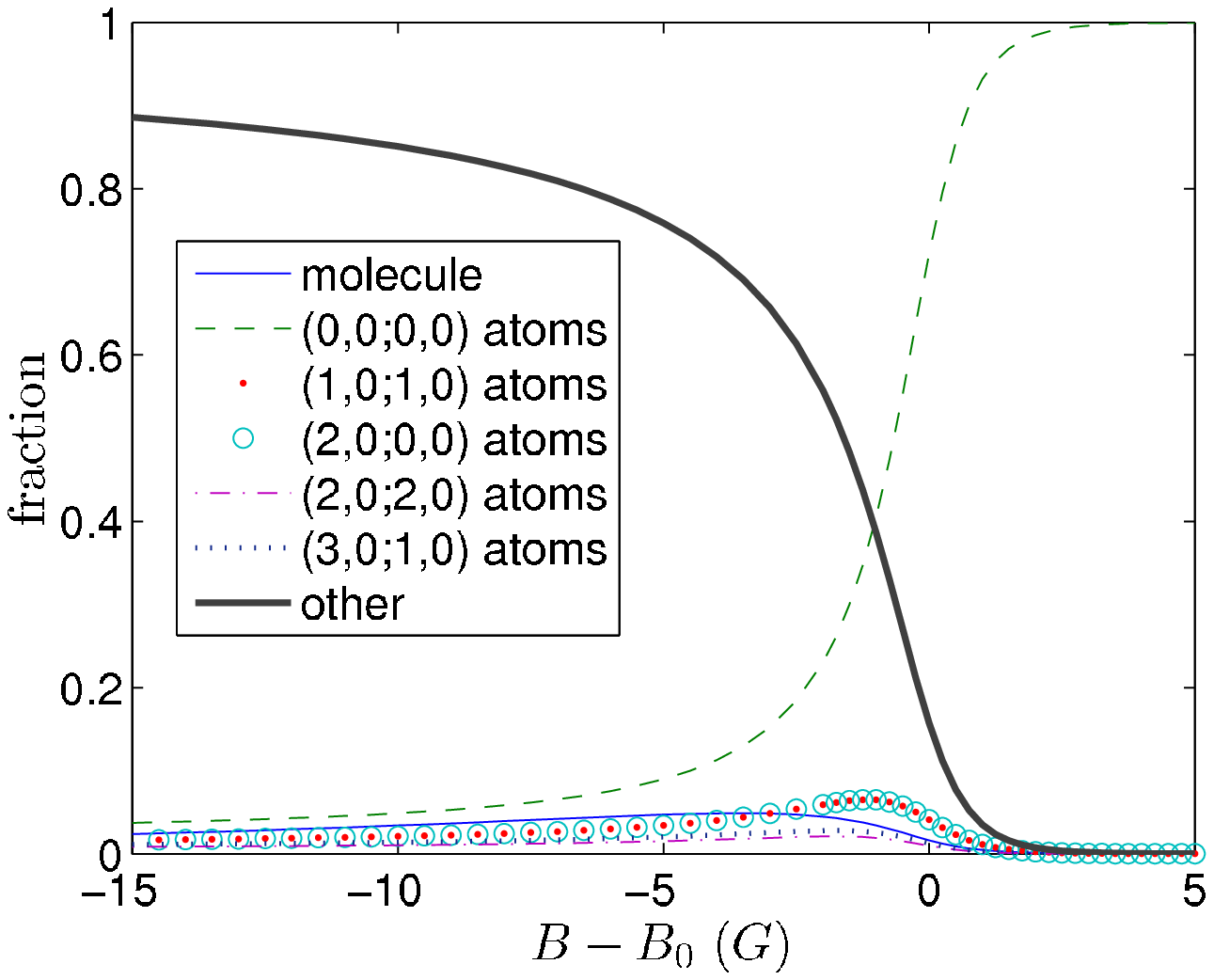}}
\subfigure[2D free $^6$Li]
        {\includegraphics[width=.73\columnwidth]{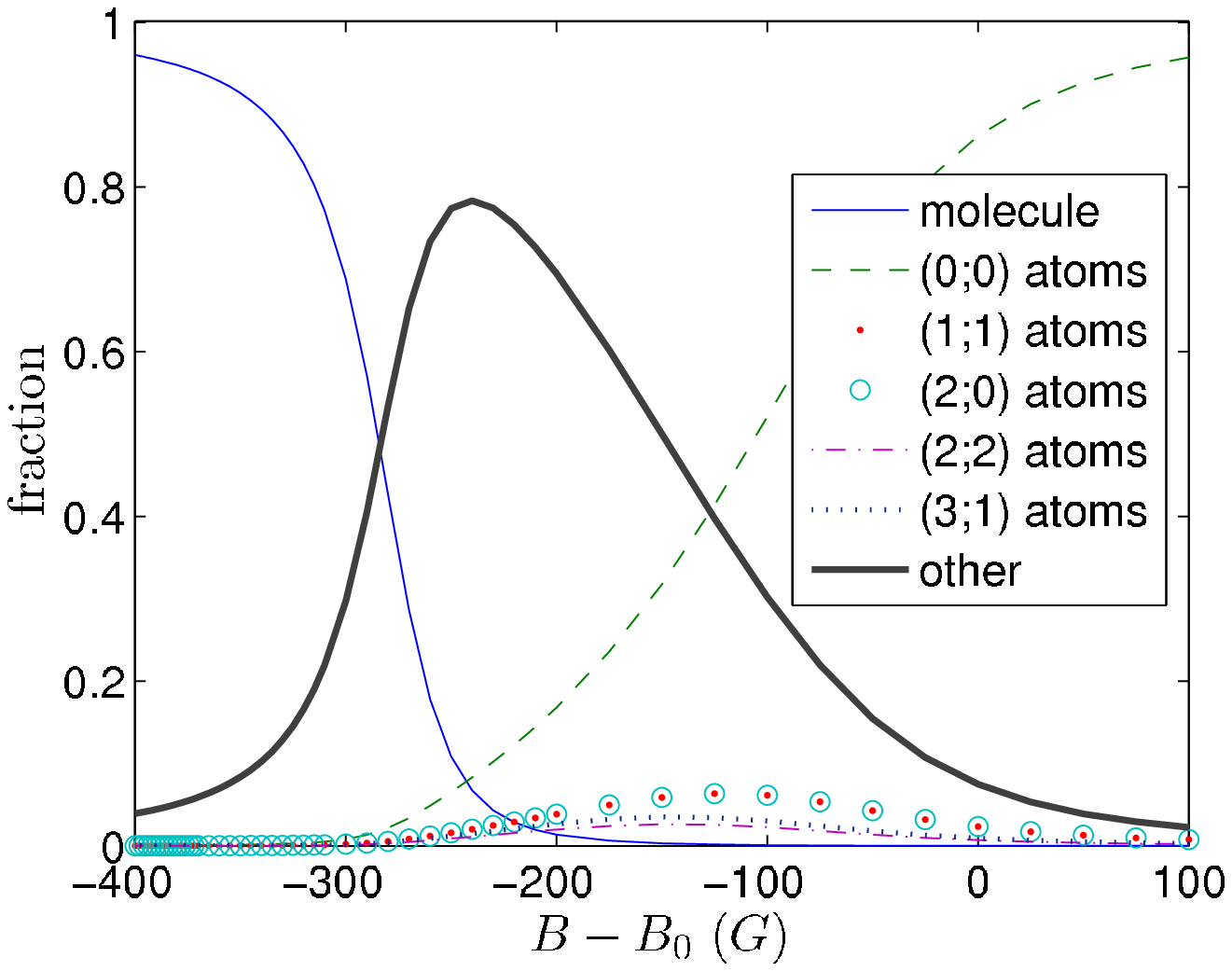}}
\subfigure[1D free $^6$Li]
        {\includegraphics[width=.73\columnwidth]{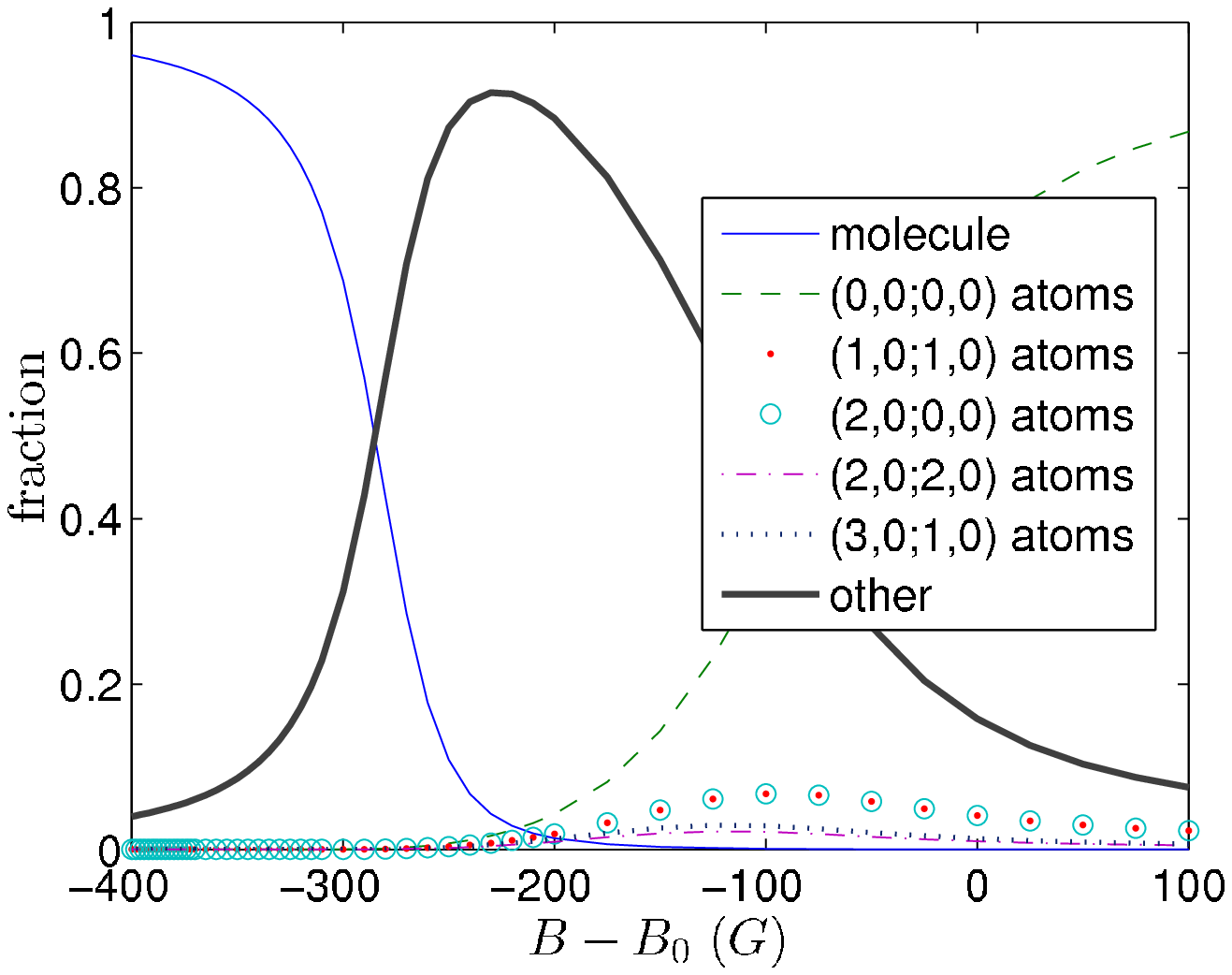}}
\caption{Ground state composition vs. detuning. Only the six most
significant components are shown explicitly. The label $\left(
\mathbf{m};\mathbf{n}\right) $ denotes the sum of
$P_{\mathbf{mn}}=\sum_{\mathbf{k}}\protect\eta _{\mathbf{mnk}}^{2}$
and all components identical by symmetry.} \label{fig:gspop}
\end{figure*}

To answer this question, we use the above formalism for the two-atom
state, and calculate the population distribution in the transverse
levels for $^{40} $K and $^{6}$Li atoms, which are the relevant
species for the current experiments. We take the scattering
parameters $W\simeq 8\,G$, $a_{bg}\simeq 174\,a_{B},$ $\mu
_{co}\simeq 1.68\mu _{B}$ for $^{40}$K \cite{9} and $W\simeq
300\,G$, $a_{bg}\simeq -1405\,a_{B}$, $\mu _{co}\simeq 2\mu _{B}$ for $^{6}$%
Li \cite{17}. With a typical trap frequency $\omega \simeq 2\pi \times
62\,kHz$ \cite{3}, the physical parameters for the atom-molecule interaction
are then given by $g_{p}=23$ ($272$), $U_{p}=1.7$ ($-5.5$) for $^{40}$K ($%
^{6}$Li). To calculate the transverse population distribution in the
system ground state, we first determine the binding energy between
the atoms using Eq. \eqref{eq:energy}. The binding energy
$E_{B}=E-\left( 3-d\right) /2$ (in the unit of $\hbar \omega $),
where the latter term $\left( 3-d\right) /2 $ is the energy for two
free atoms (excluding the contribution from the center-of-mass
mode). As the binding energy is of interest by itself, we show
$E_{B}$ as a function of the magnetic field detuning $B-B_{0}$ in
Fig. \ref{fig:EvsB_K} and \ref{fig:EvsB_L} with $d=1,2$ for $^{40}$K
and $^{6}$Li, respectively.
For $^{40}$K, the binding energy saturates on the deep BEC side.
This effect comes from the positive background scattering length
of $^{40}$K, and is related to the weakly bound state in the open
collision channel. The detailed discussion of the background
scattering effects can be found in Ref. \cite{18} (see also
\cite{19}). So, outside of the near resonance region, the result
here is quite different from the one in Ref. \cite{13}, where the
binding energy is calculated for the $d=1$ case without the
background scattering contribution. From Fig. \ref{fig:EvsB}, one
can also see that with a transverse trap, the binding energy
$\left| E_{B}\right| $ is always positive, and only tends to zero
when one goes to the deep BCS limit \cite{3,7,13} (or when
$B-B_{0}\rightarrow W$ for $^{40}$K, where the scattering length
goes to zero). This is distinct from the case without traps, where
$E_{B}=0 $ for $B\geq B_{0}$.

\begin{figure}[tbp]
\subfigure[$^{40}$K, $B-B_{0}=-18\;G$]
        {\includegraphics[width=.75\columnwidth]
        {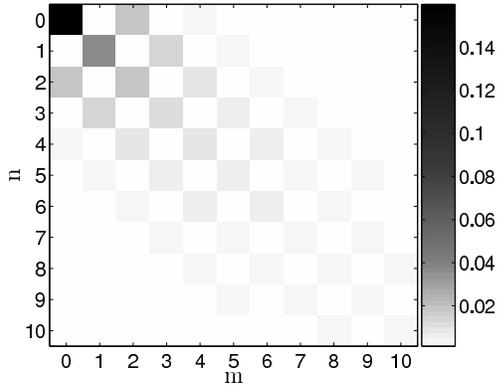}}
\subfigure[$^6$Li, $B-B_{0}=-200\;G$]
        {\includegraphics[width=.75\columnwidth]
        {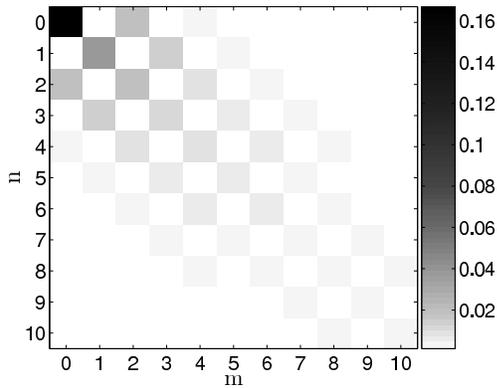}}
\caption{The matrix $\sum_{\mathbf{k}}\protect\eta
_{mn\mathbf{k}}^{2}$ for 2D free atoms on the BEC side.}
\label{fig:eta_mnk}
\end{figure}

In Fig. \ref{fig:gspop}, we show the population distribution in
transverse levels for $^{40}$K and $^{6}$Li in two or one
dimensional traps as a function of the magnetic field detuning. The
population fraction in the transverse modes $\left(
\mathbf{m;n}\right) $\ is defined as
$P_{\mathbf{mn}}=\sum_{\mathbf{k}}\left| \eta_{\mathbf{mnk}}\right|
^{2}$. In Fig. \ref{fig:eta_mnk}, we show a more complete picture of
the population distribution at a fixed detuning on the BEC side for
atoms in a one dimensional trap.
From the figures, one can see that in general many transverse modes
are populated. For a fixed mode, the population still goes down as
the energy of the mode goes up, but there are so many excited
transverse modes that the total population fraction in the excited
levels actually dominates in typical configurations.

\begin{figure}[tbp]
\subfigure[2D free $^{40}$K]
        {\includegraphics[height=.48\columnwidth]
        {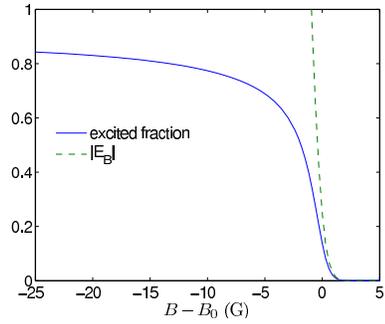}}
\subfigure[1D free $^{40}$K]
        {\includegraphics[height=.48\columnwidth]
        {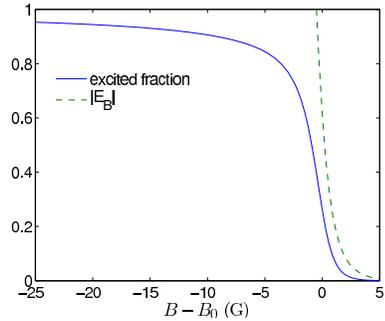}}
\subfigure[2D free $^6$Li]
        {\includegraphics[height=.48\columnwidth]
        {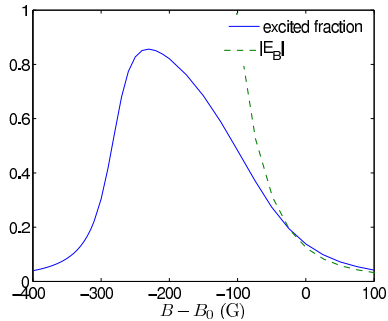}}
\subfigure[1D free $^6$Li]
        {\includegraphics[height=.48\columnwidth]
        {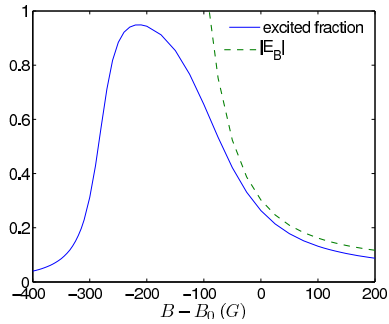}}
\caption{Excited fraction and binding energy vs. detuning for
$\protect\omega \simeq 2\protect\pi \times 62\,kHz$.}
\label{fig:E+exfrac}
\end{figure}

In Fig. \ref{fig:E+exfrac}, we draw the overall fraction populating
the transverse
excited levels, which is defined as $P_{ex}\equiv 1-\beta ^{2}-\sum_{\mathbf{%
k}}\eta _{\mathbf{00k}}^{2}$. This fraction needs to satisfy
$P_{ex}\ll 1$ for the assumption that the atoms only populate the
lowest transverse level. From the figure, one can see that this
condition is in general not satisfied for $^{40}$K and $^{6}$Li,
except in the deep BCS limit with the binding energy $|E_{B}|\ll 1$.
\begin{figure*}[tbp]
\subfigure[2D free with $U_p=+1.7$]
        {\includegraphics[width=.85\columnwidth]
        {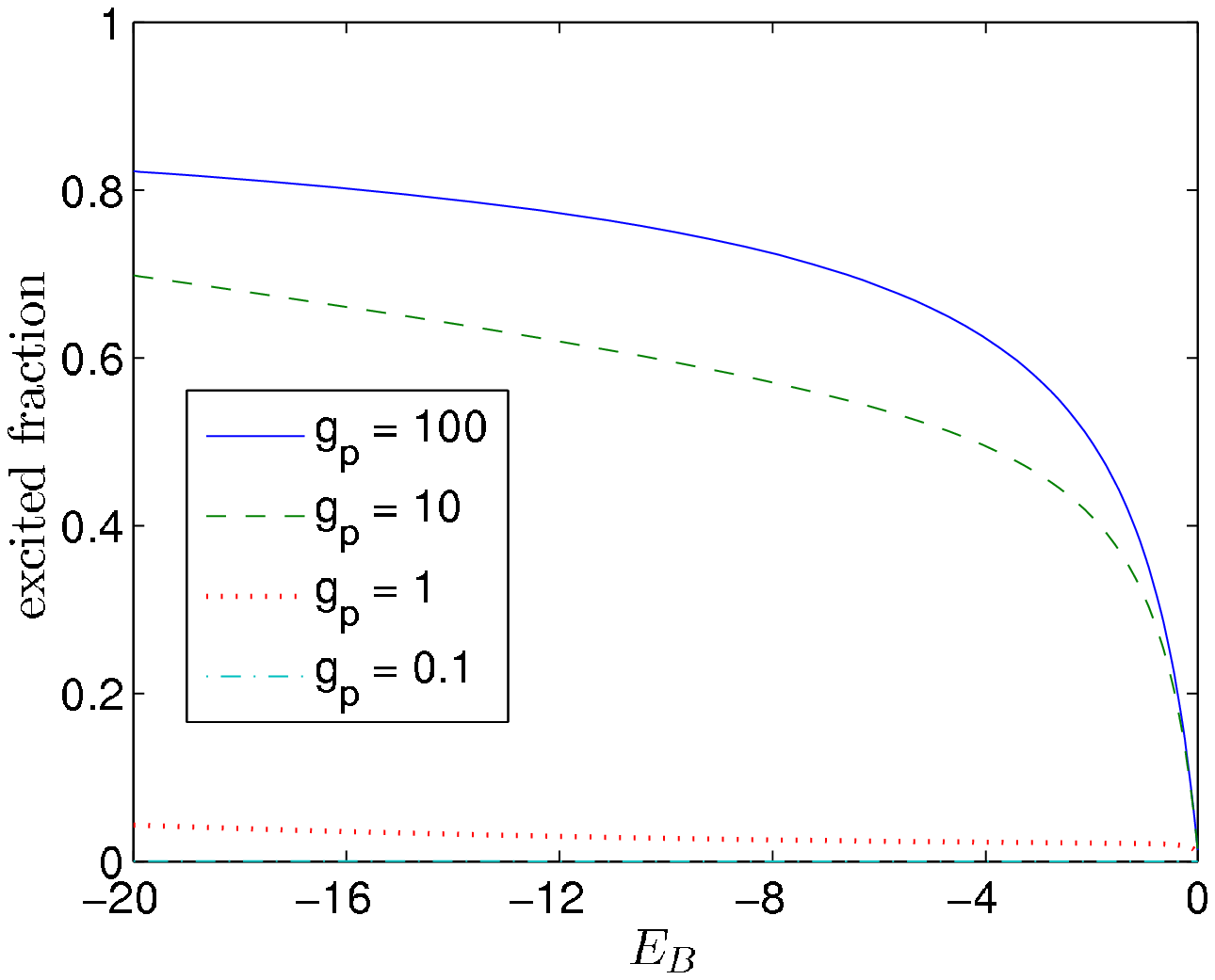}}
\subfigure[1D free with $U_p=+1.7$]
        {\includegraphics[width=.85\columnwidth]
        {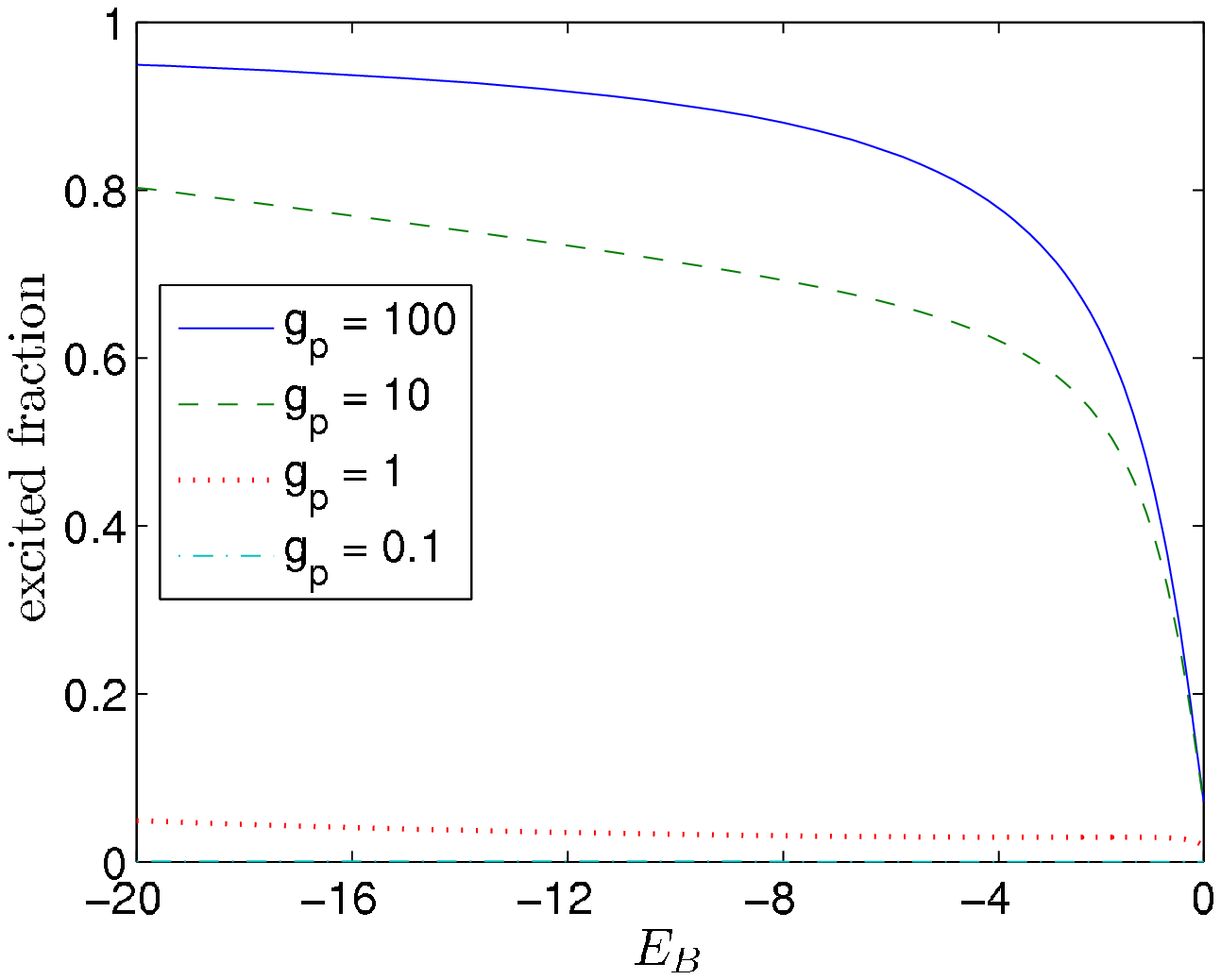}}
\subfigure[2D free with $U_p=-5.5$]
        {\includegraphics[width=.85\columnwidth]
        {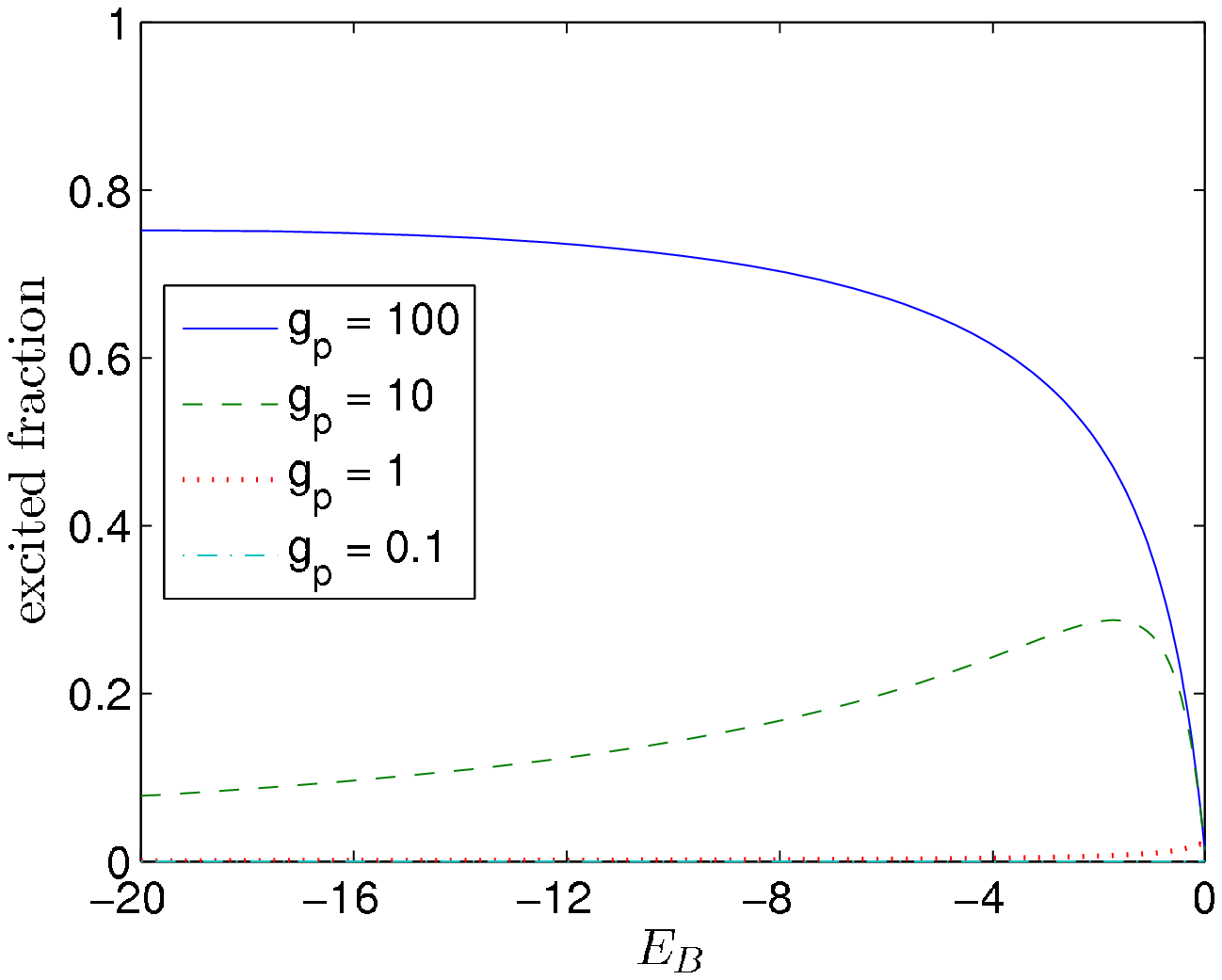}}
\subfigure[1D free with $U_p=-5.5$]
        {\includegraphics[width=.85\columnwidth]
        {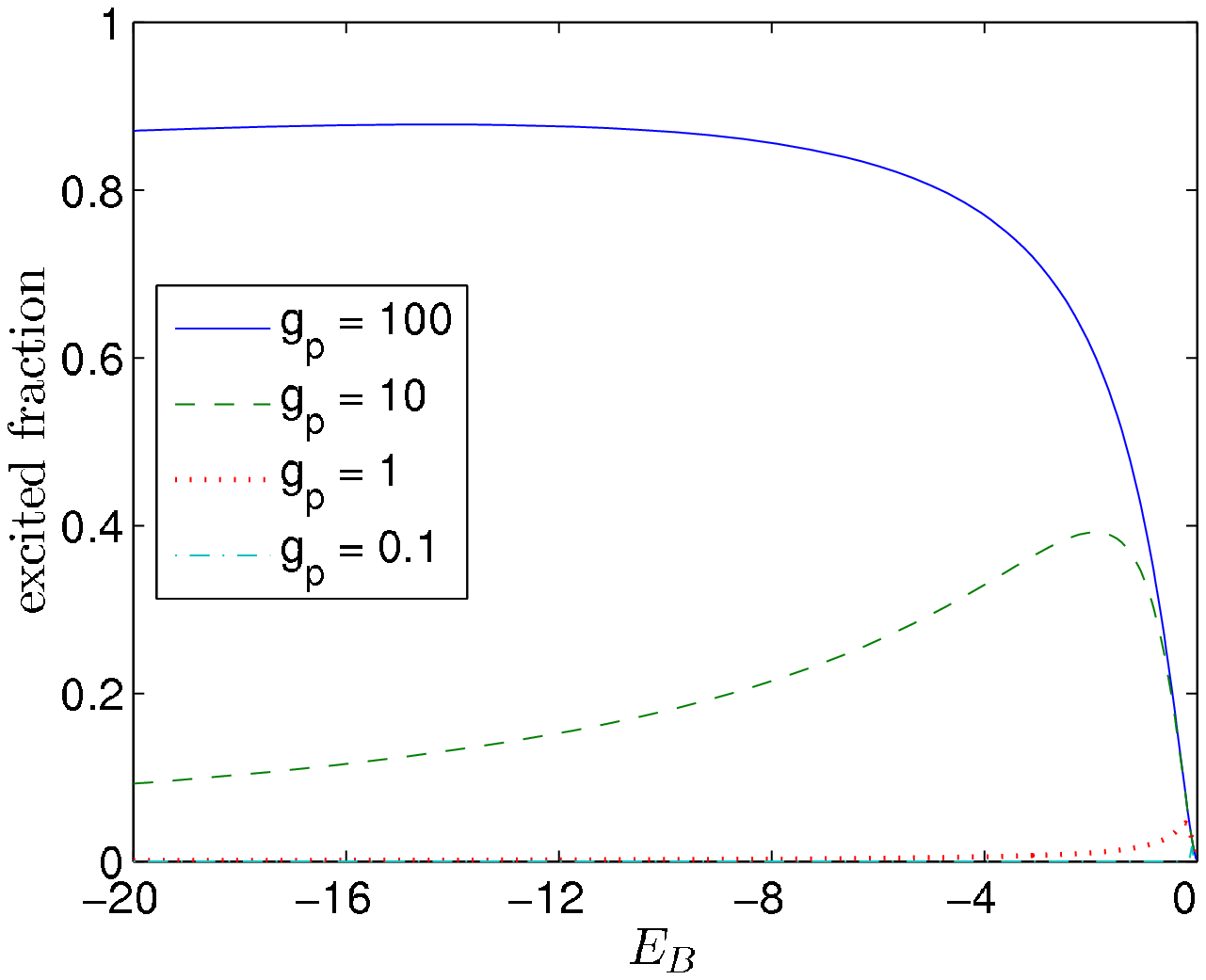}}
\caption{Excited fraction vs. binding energy for various coupling
rates.}\label{fig:exfracvsg}
\end{figure*}
For the case of $^{6}$Li, if one goes to the very deep BEC limit
with the closed channel population (the bare molecule fraction)
$\beta \rightarrow 1$, the condition $P_{ex}\ll 1$ is also
automatically satisfied. For $^{40}$K, because of its positive
background scattering length, the bare molecule fraction $\beta $
remains small even if one goes to the deep BEC limit \cite{18,19}.
So the excited fraction $P_{ex}$ continuously goes up as one
increases the negative detuning. From this calculation, it is clear
that for the entire region near resonance, which is of particular
experimental interest, one cannot neglect the transverse excited
fractions for any atoms with a wide Feshbach resonance.

The condition $P_{ex}\ll 1$\ can only be satisfied in the near
resonance region for the narrow Feshbach resonance. In Fig.
\ref{fig:exfracvsg}, we show the excited fractions for various
atom-molecule coupling rates. The background scattering length still
takes the positive or negative values corresponding to $^{40}$K or
$^{6}$Li atoms, but we vary the coupling rate $g_{p}$ (so the
resonance width $W$ changes). One can see that the condition
$P_{ex}\ll 1$
is satisfied in the whole region only when $g_{p}<1$. As $g_{p}\propto \sqrt{%
W}$, the condition $g_{p}<1$\ requires a very narrow resonance
with the resonance width $W<0.01G$. On the other hand, one can
also see from the figure that to satisfy $P_{ex}\ll 1$, the
background interaction $\left|U_{p}\right|$ can be somewhat larger
than $1$, but it cannot be arbitrarily larger. We have tested (not
shown in the figure) that the condition $P_{ex}\ll 1$ breaks down
when $U_{p}$ is on the order of a few tens.

\begin{figure*}[tbp]
\subfigure[2D free $^{40}$K]
        {\label{fig:exfracK-a}\includegraphics[width=.85\columnwidth]
        {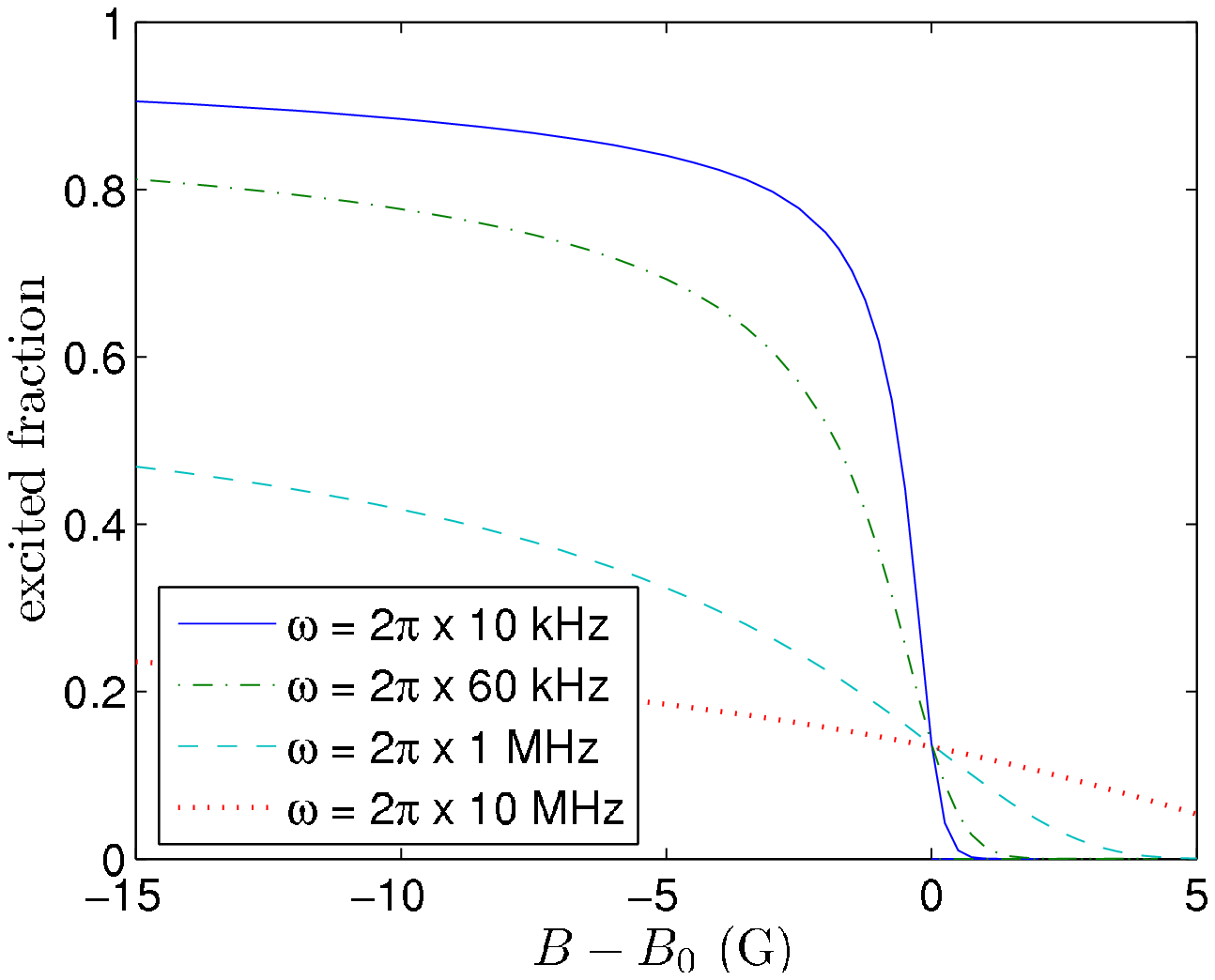}}
\subfigure[1D free $^{40}$K]
        {\label{fig:exfracK-b}\includegraphics[width=.85\columnwidth]
        {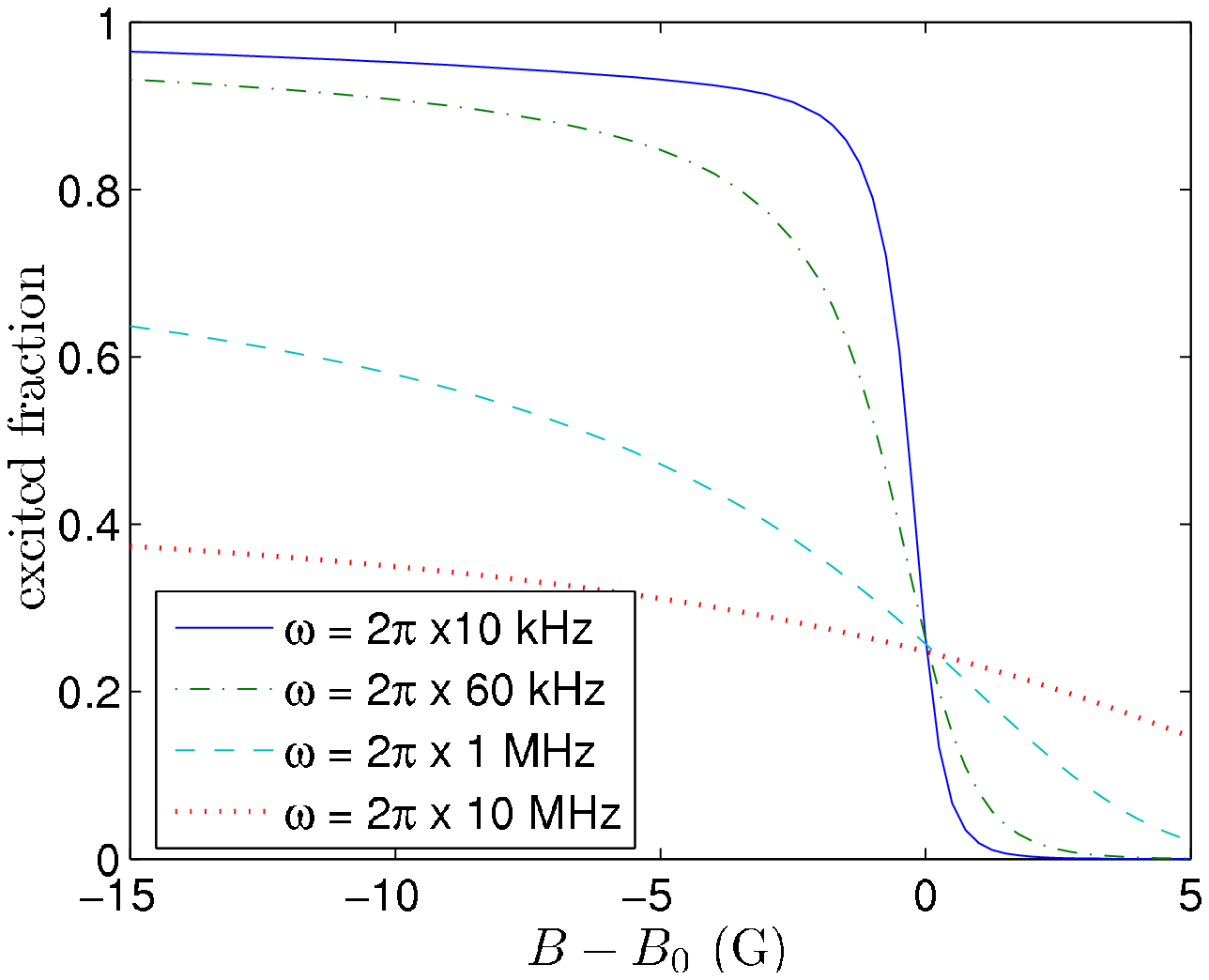}}
\subfigure[2D free $^6$Li]
        {\label{fig:exfracL-a}\includegraphics[width=.85\columnwidth]
        {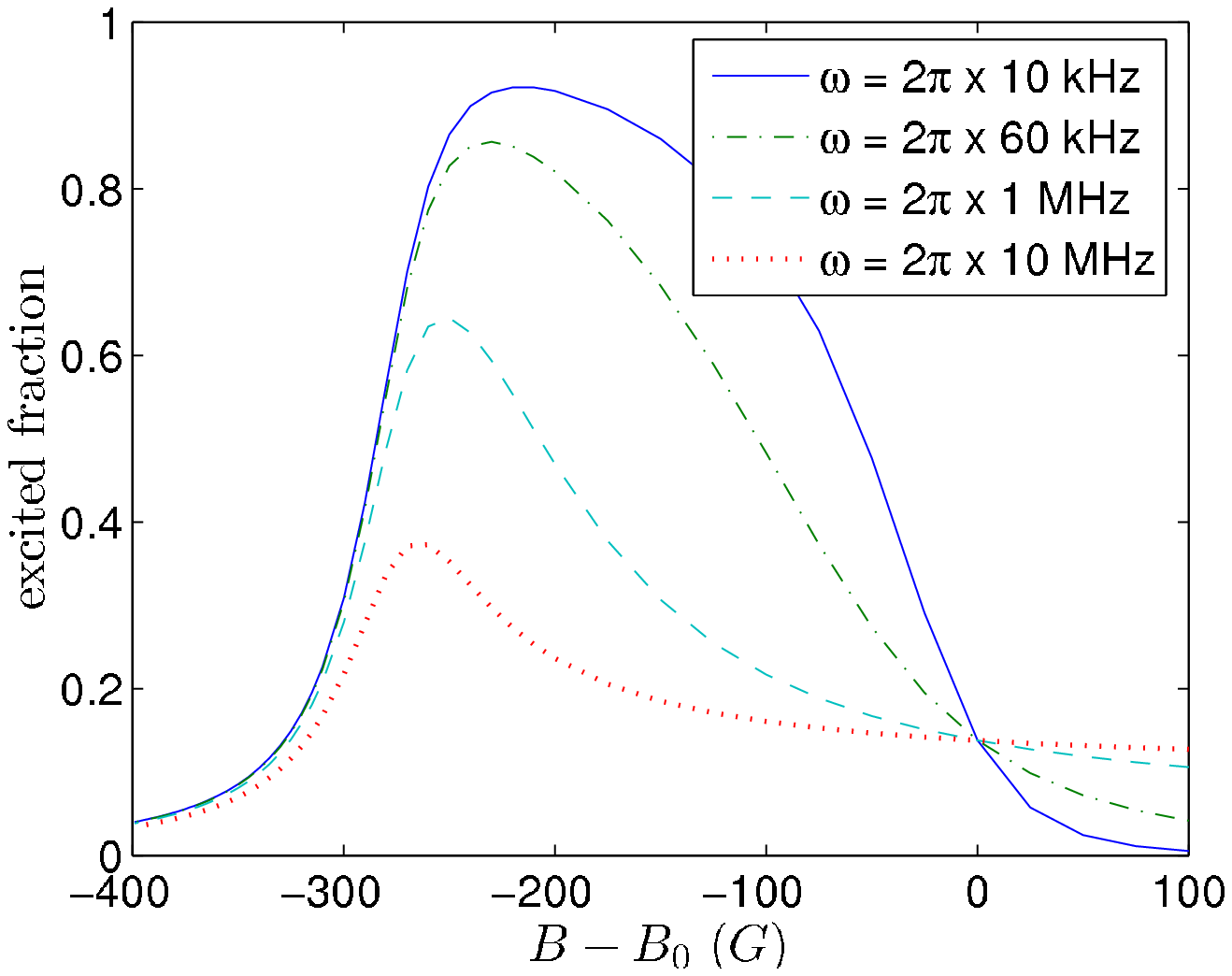}}
\subfigure[1D free $^6$Li]
        {\label{fig:exfracL-b}\includegraphics[width=.85\columnwidth]
        {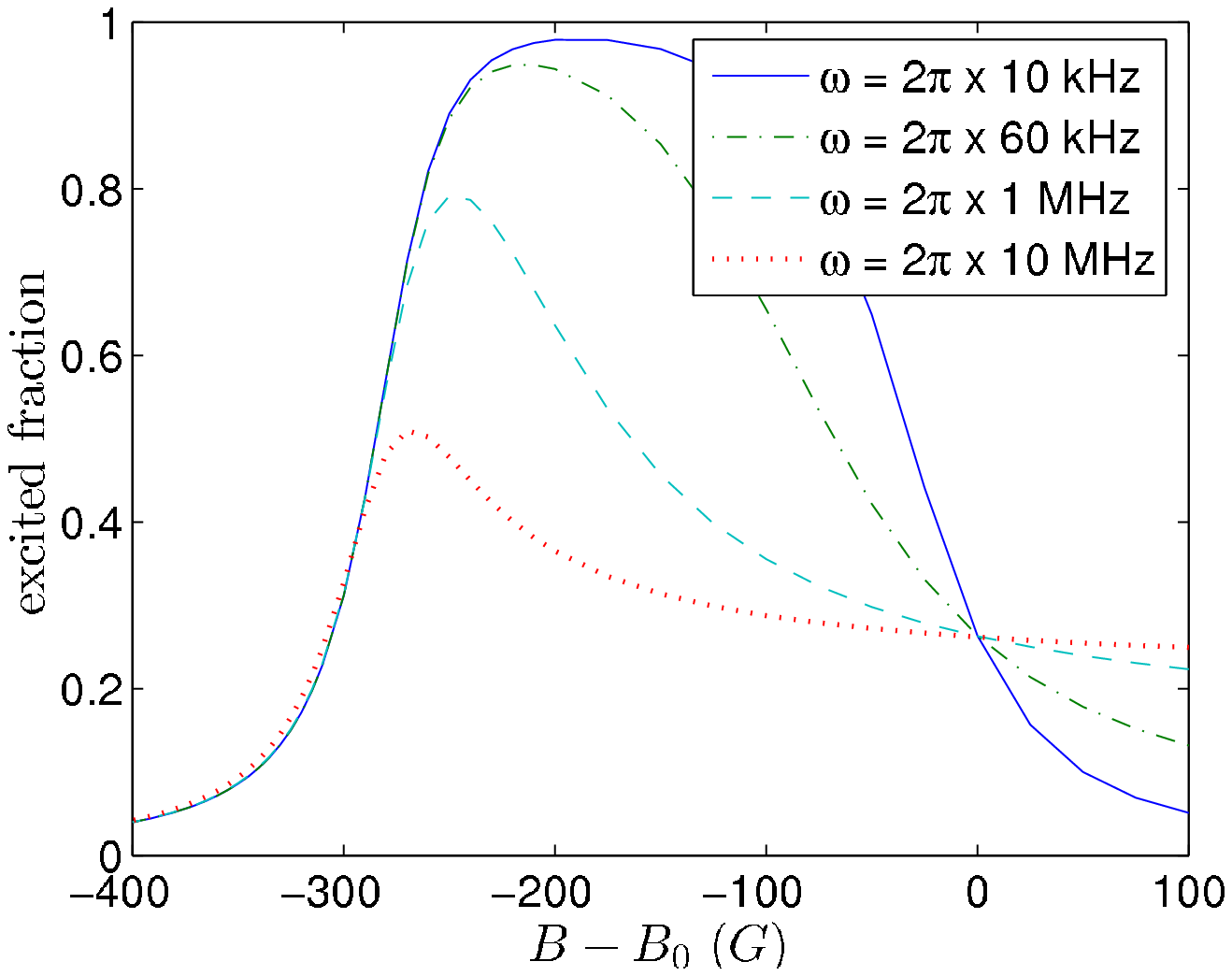}}
\caption{Excited fraction vs. detuning for various trapping
frequencies.} \label{fig:exfrac}
\end{figure*}

The above calculations are done with a fixed trap frequency $\omega
\simeq 2\pi \times 62\,kHz$, as it is typical for current
experiments \cite{3}. One may expect that if the trap frequency
further increases, it will become much easier to suppress the
transverse excited fraction and to satisfy the condition $P_{ex}\ll
1$. To test whether this is true, we show in Fig. \ref {fig:exfrac}
the transverse excited fraction versus the magnetic field detuning
for various magnitudes of the trap frequency.
One can see that even if the trap frequency increases by several
orders of magnitudes, the transverse excited fraction $P_{ex}$ does
not change much and remains significant. To understand this puzzling
effect, we note that although the transverse level spacing increases
a lot, the effective atom-molecule coupling rate also increases. One
cannot assume a constant atom-molecule coupling rate as an enhanced
trapping in the transverse direction will suppress the pair size in
all directions including the free dimensions (see the next
paragraph). The effective atom-molecule coupling rate increases
significantly when the pair size shrinks. As a net effect, the ratio
between the atom-molecule coupling and the transverse level spacing
is only a slowly-varying function of the trap frequency. The
dimensionless parameter $g_{p}$, which measures the effective
atom-molecule coupling strength in the unit of the trap frequency,
actually characterizes this ratio. From its expression, one can
verify that $g_{p}$ goes as $\omega ^{-1/4}$. So, the effective
ratio $g_{p}$ only drops by a factor of $10$ (it will still be
larger than $1$ for $^{6}$Li and $^{40}$K) even if the trap
frequency increases by four orders of magnitudes from its current
value (which is almost impossible). This explains the relative
insensitivity of the excited fraction $P_{ex}$ to the trap
frequency. We can also conclude here that for any reasonable trap
strength one can imagine, it is impossible
to neglect the transverse excited fractions for realistic atoms such as $%
^{6} $Li and $^{40}$K with a wide Feshbach resonance.

With an increased transverse trapping, it is easy to understand that
the pair size along that direction will be suppressed. However, it
is not so obvious that the pair size along the free dimension(s),
where there is no trap, will correspondingly shrink. The latter
actually comes from the interaction effect. Under strong
interaction, the pair size cannot change only along one direction.
To see the pair size shrinking along the untrapped directions, we
can take the Fourier transform of the pair wavefunction $\eta
_{\mathbf{mnk}}$ with respect to the momentum $\mathbf{k}$
in the free dimension(s). This Fourier transform, denoted as $\eta _{\mathbf{%
mn}r}$, yields the real-space wavefunction, which is the
probability amplitude for an atom pair in transverse modes
$\mathbf{m}$ and $\mathbf{n}$ to be spaced a distance $r$ apart in
free dimension(s).
\begin{figure*}[ht]
\subfigure[2D free $^{40}$K, $B-B_{0}=-2\,G$]
        {\label{fig:xiK-a}\includegraphics[height=.6\columnwidth]
        {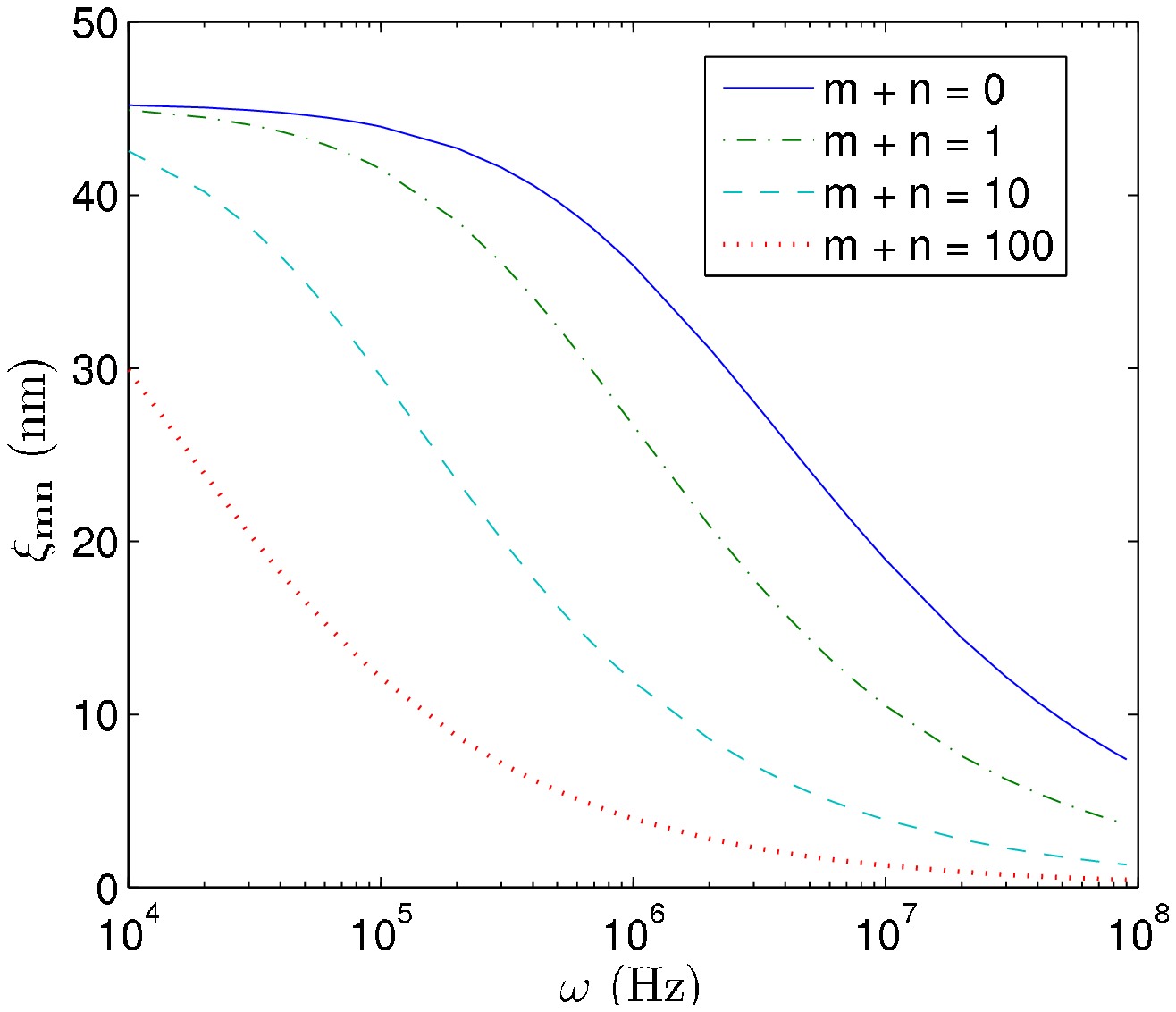}}
\subfigure[1D free $^{40}$K, $B-B_{0}=-2\,G$]
        {\label{fig:xiK-b}\includegraphics[height=.6\columnwidth]
        {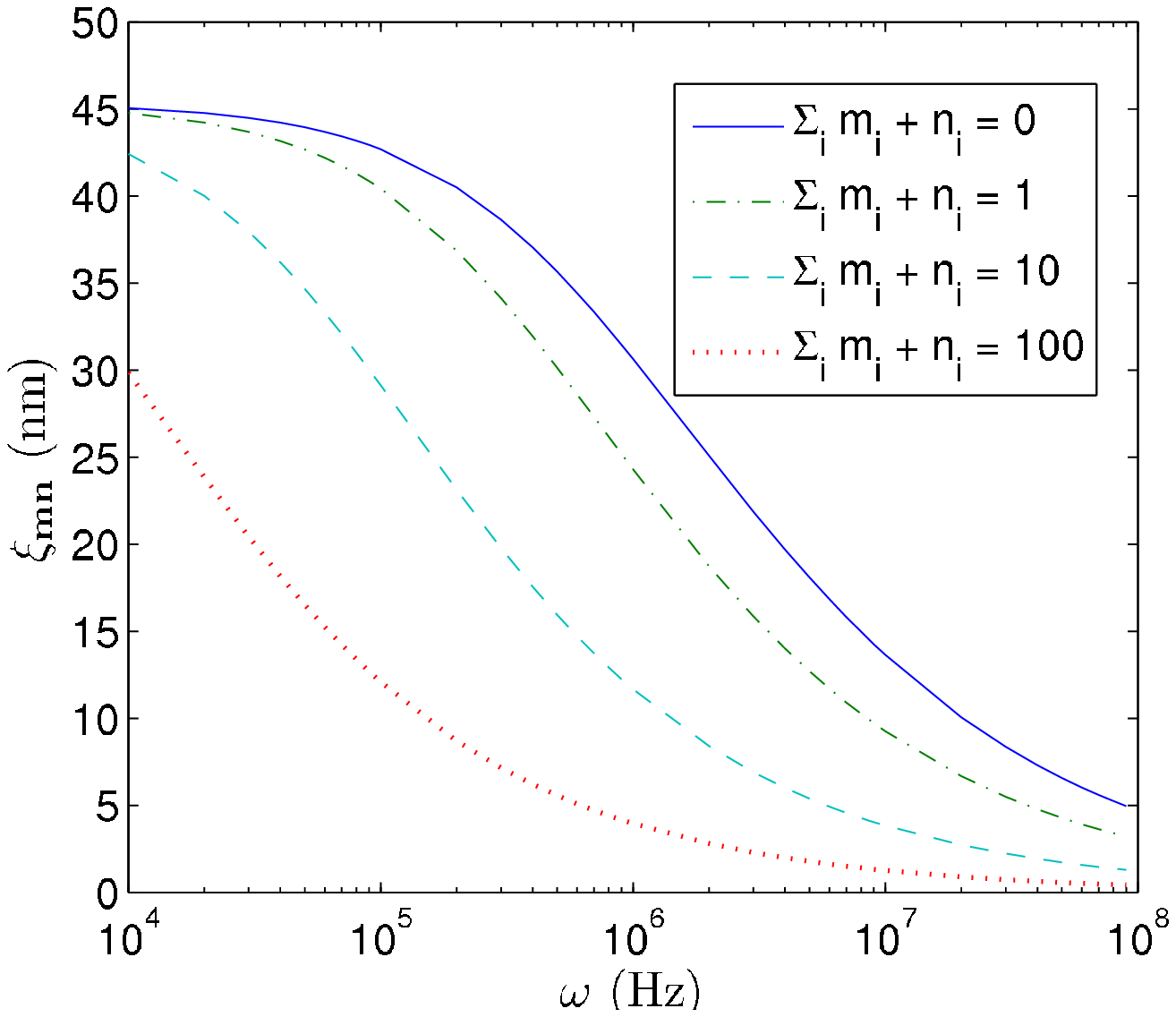}}
\subfigure[2D free $^6$Li, $B-B_{0}=-200\,G$]
        {\label{fig:xiL-a}\includegraphics[height=.6\columnwidth]
        {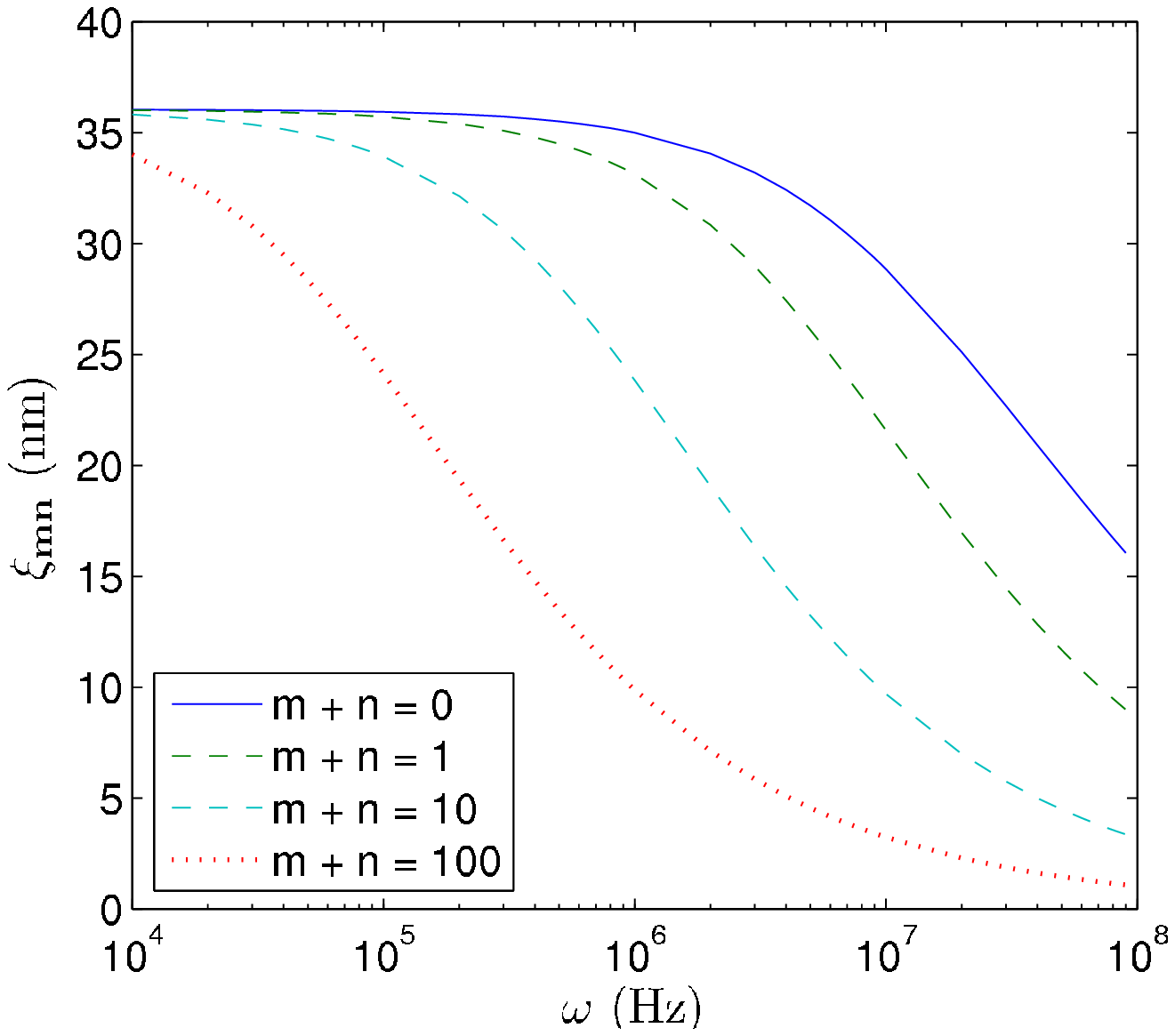}}
\subfigure[1D free $^6$Li, $B-B_{0}=-200\,G$]
        {\label{fig:xiL-b}\includegraphics[height=.6\columnwidth]
        {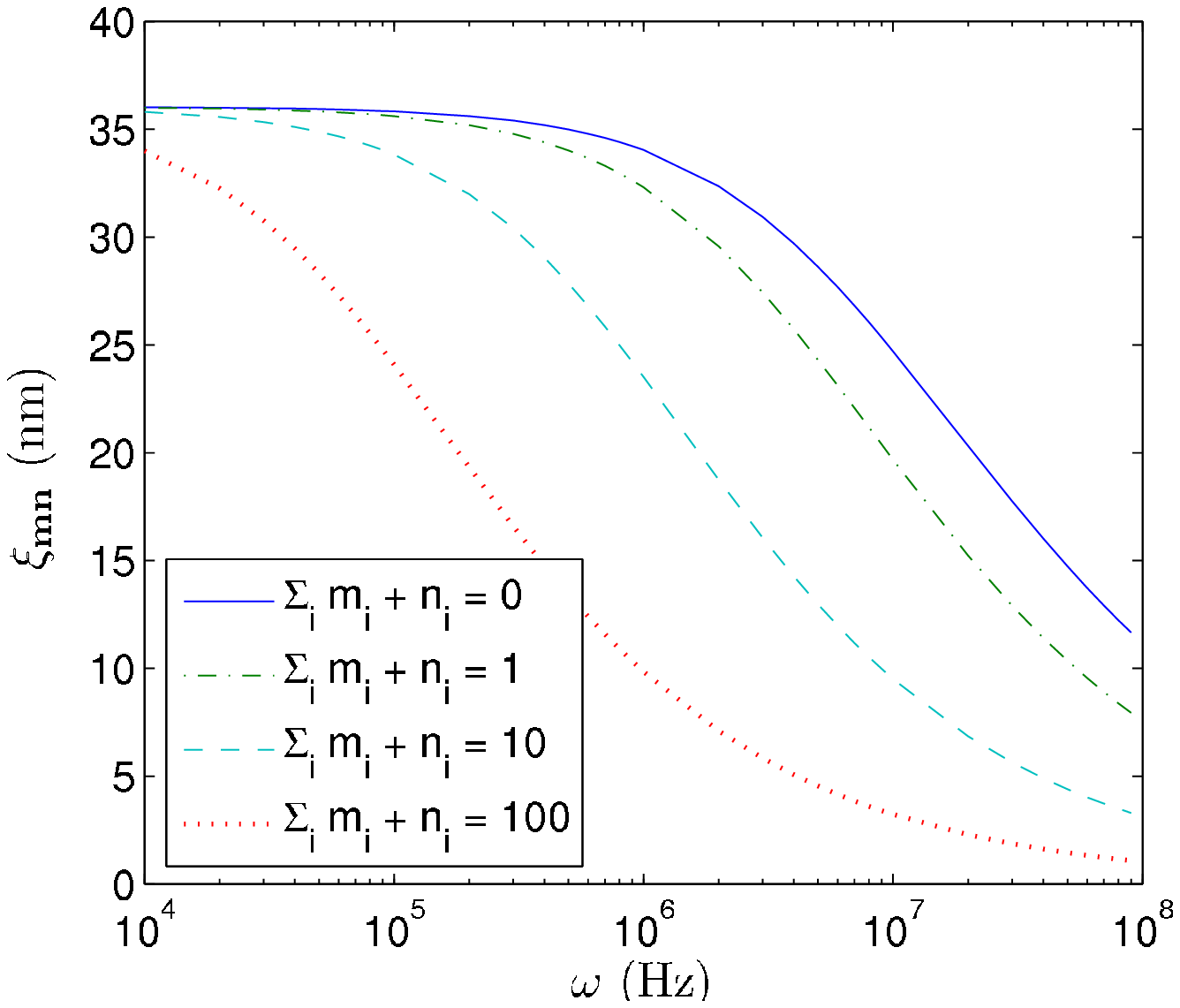}}
\caption{Characteristic atomic pair size vs. trapping frequency for
a fixed detuning.}\label{fig:xi}
\end{figure*}
From Eq. \eqref{eq:fermion}, $\eta _{\mathbf{mn}r}$ is given by
\begin{equation}
\eta _{\mathbf{mn}r}=
\begin{cases}
-\frac{g_{p}}{2\pi }\left( a_{t}L\right) ^{-1}\beta \gamma _{mn}K_{0}\left(
|r|/\xi _{mn}\right) & d=2 \\
-\frac{g_{p}}{2}\frac{\xi _{\mathbf{mn}}}{\sqrt{a_{t}^{3}L}}\beta \gamma _{%
\mathbf{mn}}e^{-|r|/\xi _{\mathbf{mn}}} & d=1
\end{cases}
\;,
\end{equation}
where $K_{0}\left( x\right) =\int_{0}^{\infty }dt\cos \left( xt\right) /%
\sqrt{t^{2}+1}$ is a modified Bessel function of the second kind, and
\begin{equation}
\xi _{\mathbf{mn}}=\frac{a_{t}}{\sqrt{-E_B+\sum_{i=1}^{3-d}%
\left( m_{i}+n_{i}\right) }}\;,
\end{equation}
which characterizes the pair size in the free dimension(s). The
dependence of $\xi _{\mathbf{mn}}$ on the trapping frequency is
shown in Fig. \ref{fig:xi} for different transverse modes $\left(
\mathbf{mn}\right) $ on the BEC side. Note that under typical
configurations, the atom population is broadly distributed over the
transverse modes (as illustrated by Fig. \ref{fig:eta_mnk}), so the
exact pair size in the free dimension(s) should come from the
average of $\xi _{\mathbf{mn}}$\ over different $\left(
\mathbf{mn}\right) $.
From Fig. \ref{fig:xi}, one can see that for typical values of
$\left( \mathbf{mn}\right) $, the pair size significantly shrinks
with increase of the trap frequency, so its average will follow
the same trend. As one moves towards the BCS side of resonance,
the trend only becomes more dramatic. This provides a physical
interpretation of the coupling enhancement: The transverse trap
raises the minimum energy of the atoms relative to the molecules.
Thus, the smaller atom pairs which have stronger mixture with the
molecules become more energetically favorable. The net result is a
sort of induced pair-wise confinement along the free dimension(s),
and this increased local density causes the effective coupling to
increase.

\section{Summary}

In summary, we have shown that for experimentally relevant cases,
atoms trapped along one or two dimensions cannot be considered to
be in the transverse ground level except when well out of the
BCS-BEC crossover region. In the crossover region and for the
ground state of the system, a significant fraction of atomic
population resides in the excited transverse levels, even if the
gas is very dilute. Furthermore, one cannot effectively suppress
the transverse excited fraction by raising the trap intensity.
Even with an extremely strong trap far beyond the current
experimental technology, the transverse excited fraction is not
negligible yet for realistic atoms such as $^{6}$Li and $^{40}$K
across a wide Feshbach resonance. The conclusion here is that in
the experimentally interesting region, one cannot describe
strongly interacting atoms under a transverse trap as a
low-dimensional system by assuming a fixed transverse mode.
Although this result does not exclude an \textit{effective}
low-dimensional description of this strongly interacting system,
it indeed shows that the effective description will become much
more subtle, and such a description needs to take into account the
broad population distribution of the atoms in all the transverse
modes. The derivation of such an effective description will be a
topic for further investigation \cite{12}.

This work was supported by the NSF award (0431476), the ARDA under
ARO contracts, the A. P. Sloan Foundation.


\begin{thebibliography}{99}
\bibitem{1}  B. Paredes, A. Widera, V. Murg, O. Mandel, S. Folling, I.
Cirac, G. Shlyapnikov, T. Hansch, and I. Bloch, Nature 429, 277 (2004).

\bibitem{2}  T. Kinoshita, T. Wenger, and D. Weiss, Science 305, 5687 (2004).

\bibitem{3}  M. Kohl, T. Stoeferle, K. Guenter, M. Koehl, Tilman Esslinger,
Phys. Rev. Lett. \textbf{94}, 080403 (2005); T. Stoeferle, H.
Moritz, K. Guenter, Michael Koehl, T. Esslinger, Phys. Rev. Lett.
96, 030401 (2006).

\bibitem{4}  For a theoretical review, see Y. Castin, cond-mat/0407118.

\bibitem{5}  D. E. Sheehy, Leo Radzihovsky, Phys. Rev. Lett. 95, 130401
(2005).

\bibitem{6}  E. Orignac, R. Citro, cond-mat/0601269.

\bibitem{7}  M. Olshanii, Phys. Rev. Lett. 81, 938 (1998); T. Busch, B.-G.
Englert, K. Rzazewski, M. Wilkens, Found. Physics 28, 549 (1998); D.S.
Petrov, M. Holzmann, G.V. Shlyapnikov, Phys. Rev. Lett 84, 2551 (2000); P.
O. Fedichev, M. J. Bijlsma, P. Zoller, Phys. Rev. Lett. 92, 080401 (2004).

\bibitem{8}  M. Greiner, \textit{et al.}, Nature \textbf{415}, 39 (2002); C.
Orzel, \textit{et al.}, Science \textbf{291}, 2386 (2001); D. Jaksch, et
al., Phys. Rev. Lett. \textbf{81}, 3108 (1998).

\bibitem{9}  C.A. Regal, M. Greiner and D.S. Jin, Phys. Rev. Lett. \textbf{92%
}, 040403 (2004); M.W. Zwierlein \textit{et al.}, Phys. Rev. Lett. \textbf{92%
}, 120403 (2004); C. Chin \textit{et al.}, Science \textbf{305}, 1128 (2004).

\bibitem{10}  L.-M. Duan, Phys. Rev. Lett. \textbf{95}, 243202 (2005).

\bibitem{11}  R. B. Diener, T.-L. Ho, Phys. Rev. Lett. \textbf{96}, 010402
(2006).

\bibitem{12}  J. P. Kestner, L.-M. Duan, in preparation.

\bibitem{13}  D. B. M. Dickerscheid, H. T. C. Stoof, cond-mat/0506530; D. B.
M. Dickerscheid \textit{et al.}, Phys. Rev. A \textbf{71}, 043604 (2005).

\bibitem{14}  M. Holland \textit{et al.}, Phys. Rev. Lett. \textbf{87},
120406 (2001); E. Timmermans, \textit{et al.}, Phys. Rep. \textbf{315}, 199
(1999).

\bibitem{15}  Q. Chen, J. Stajic, S. Tan, K. Levin, Physics Reports 412, 1
(2005).

\bibitem{Note}  We will let this energy cutoff $E_{c}$ tend to infinity in
our final equations as the low-energy physics there is independent
of the value of the high energy cutoff (it cancels with the
divergence of $S\left( E\right) $ in Eq. (13)).

\bibitem{17}  M. Bartenstein \textit{et al.}, Phys. Rev. Lett. \textbf{94},
103201 (2005).

\bibitem{18}  W. Yi, L.-M. Duan, cond-mat/0603264.

\bibitem{19}  M. H. Szymanska, K. Goral, T. Koehler, K. Burnett, Phys. Rev.
A 72, 013610 (2005).

\end{thebibliography}
\end{document}